\documentclass[twocolumn,showpacs,preprintnumbers,superscriptaddress,amsmath,amssymb,prl]{revtex4}

\usepackage{graphicx}
\usepackage{dcolumn}
\usepackage{bm}
\usepackage{amsmath}
\usepackage{indentfirst}

\newcommand{\be}{\begin{equation}}
\newcommand{\ee}{\end{equation}}
\newcommand{\ba}{\begin{eqnarray}}
\newcommand{\ea}{\end{eqnarray}}

\begin{document}

\preprint{APS preprint}

\title{
Multifractal Scaling of Thermally-Activated Rupture Processes
}

\author{D. Sornette}
\affiliation{Institute of Geophysics and Planetary Physics,
University of California, Los Angeles, CA 90095}
\affiliation{Department of Earth and Space Sciences, University of
California, Los Angeles, CA 90095\label{ess}}
\affiliation{Laboratoire de Physique de la Mati\`ere Condens\'ee,
CNRS UMR 6622 and Universit\'e de Nice-Sophia Antipolis, 06108
Nice Cedex 2, France}
\email{sornette@moho.ess.ucla.edu}

\author{G. Ouillon}
\affiliation{Institute of Geophysics and Planetary Physics,
University of California, Los Angeles, CA 90095}
\affiliation{Laboratoire de Physique de la Mati\`ere Condens\'ee,
CNRS UMR 6622 and Universit\'e de Nice-Sophia Antipolis, 06108
Nice Cedex 2, France}
\email{ouillon@aol.com}

\date{\today}

\begin{abstract}
We propose a ``multifractal stress activation'' model combining
thermally activated rupture and long memory
stress relaxation, which predicts 
that seismic decay rates after mainshocks follow the 
Omori law $\sim 1/t^p$ with exponents $p$ linearly increasing with the 
magnitude $M_L$ of the mainshock and the inverse temperature. 
We carefully test this prediction on earthquake
sequences in the Southern California Earthquake catalog: we find 
power law relaxations of seismic sequences triggered by mainshocks with exponents
$p$ increasing with the mainshock magnitude by approximately $0.1-0.15$
for each magnitude unit increase, from $p(M_L=3) \approx 0.6$ to $p(M_L=7) \approx 1.1$,
in good agreement with the prediction of the multifractal model. 
\end{abstract}


\maketitle

Parisi and Frisch \cite{PF} and
Halsey et al. \cite{Halsey} have introduced the extended concept
of scale invariance, called multifractality, motivated by
hydrodynamic turbulence and fractal growth aggregates respectively.
Use of the multifractal spectrum as a metric to characterize
complex systems is now routinely used in many fields, including
seismology to describe the hierarchical structure in space and time
of earthquakes and faults (see for instance \cite{Godano,Main,Ouillon1}).
However, the origin of multifractality is rarely identified. This is
certainly true for earthquakes for which the possible existence of 
multifractality is under scrutiny due to limited and corrupted
data sets leading to biases \cite{Ouillon2} and its origin a matter of debate:
fractal growth processes \cite{anne}, self-organized criticality \cite{Main}
or  hierarchical cascades of stresses \cite{Rodkin} are among the physical
scenarios proposed to lead to multifractality in fault and earthquake patterns.
Here, we propose a physically-based ``multifractal stress activation''
model of earthquake interaction and triggering
based on two simple ingredients: (i) a seismic rupture results from 
thermally activated processes
giving an exponential dependence on the local stress; (ii) the stress relaxation has
a long memory. The interplay between these two physical processes 
are shown to lead to a multifractal organization of seismicity, which we
observe quantitatively in real catalogs.

Thermal activation is relevant
in all previously proposed physical processes underlying earthquakes: 
creep rupture, stress corrosion and
state-and-velocity dependent friction. 
We model seismic activity $\lambda({\vec r}, t)$ 
at position $\vec r$ and time $t$ as the occurence of
frictional sliding events and/or fault ruptures that are thermally activated processes 
facilitated by the applied stress field:
$\lambda({\vec r}, t) \sim \exp \left[-\beta E ({\vec r}, t) \right]$,
where $\beta$ is the inverse temperature and the 
energy barrier $E({\vec r}, t)$ for rupture can be written as the sum of a contribution
$E_0({\vec r})$ 
characterizing the material and of a term linearly decreasing with the locally
applied stress $\Sigma({\vec r}, t)$: $E(\vec {r}, t) = E_0(\vec {r}) - V \Sigma(\vec {r}, t)$.
$V$ is a constant which has the dimension of a volume and $\Sigma({\vec r}, t)$
is the total stress at position ${\vec r}$ and time $t$. The decrease
of the energy barrier $E({\vec r}, t)$ as a function of 
the applied stress $ \Sigma({\vec r}, t)$ embodies 
the various physical processes aiding rupture activation under stress.
In addition, there are many evidences for a stress-controlled earthquake activation process,
suggesting that earthquakes trigger earthquakes directly and indirectly via
dynamical and static stress transfers. Visco-elastic models of stress relaxation
can account for the short-term relaxation processes of the strain measured by geodetic methods
but, over long time scales, it is necessary to 
take into account the presence and geometry of lower crustal and mantle shear zones,
which lead to slower decaying relaxation rates. We thus write
the stress $\Sigma({\vec r}, t)$ at position $\vec r$ and time $t$
as the sum of contributions from
all past events at earlier times $\tau<t$ and positions ${\vec r}~'$:
$\Sigma({\vec r}, t) = \Sigma_{\rm far~field}({\vec r}, t) + 
 \int_{-\infty}^t  \int dN[d{\vec r}~' \times d\tau]
\Delta \sigma({\vec r}~', \tau)  g({\vec r} - {\vec r}~', t-\tau)$.
A given
past event at $({\vec r}~',\tau)$ contributes to the stress at
$(\vec r, t)$ by its stress drop amplitude $\Delta \sigma({\vec r}~', \tau)$
which is transfered in space and time via the stress kernel (or Green function)
$g({\vec r} - {\vec r}', t-\tau)$, taking into account both time relaxation
and spatial geometrical decay.
The term $dN[d{\vec r}~' \times d\tau]$ is the number of 
events in the volume $d{\vec r}~'$ that occurred between $\tau$ and $\tau+d\tau$.

In this letter, we restrict our analysis to the time domain. For this,
we assume for 
simplicity that $g({\vec r}, t)$ is
separable as $g({\vec r}, t) = f({\vec r}) \times h(t)$. This obtains
\be
\lambda({\vec r}, t) = \lambda_{\rm tec} ({\vec r}, t)  
~\exp \left[ \beta \int_{-\infty}^t d\tau
~s({\vec r}, \tau)  h(t-\tau) \right]~,
\label{fund2}
\ee
where $s({\vec r}, \tau) = \int d{\vec r}~' ~\Delta \sigma({\vec r}~', 
\tau) ~f({\vec r} - {\vec r}~')$
is the effective source at time $\tau$ at point ${\vec r}$ resulting from all
events occurring in the spatial domain at the same time $\tau$. 
$\lambda_{\rm tec} ({\vec r}, t)$ is the spontaneous seismicity rate in
absence of stress triggering by other earthquakes and accounts for the
tectonic loading (far field stress), which may in general be
non-homogeneous in space and perhaps depends on time. Since expression
(\ref{fund2}) is defined for any ${\vec r}$, we drop the reference to
${\vec r}$ without loss of generality. 

To go further, we specify the distribution $P(s)$ of stress sources
and the memory kernel $h(t)$.
On the basis of theoretical calculations, simulations
and measurements of rotations of earthquake focal mechanisms,
Kagan \cite{Kagan} 
has suggested that $P(s)$ should follow a 
symmetric Cauchy distribution. To capture in a phenomenological way the extended nature and
complexity of earthquake ruptures, we use a more general power law distribution
$P(s) \sim C/|\Delta s|^{1+\mu}$, 
which generalizes the Cauchy case $\mu=1$. To account for the slower-than-exponential
stress relaxation processes discussed above, we postulate that
$h(t) = {c^{1+\theta} \over (t+c)^{1+\theta}}$ for $0< t \leq T$,
which is of the Omori form with the usual small time-scale cut-off $c$. 
To ensure convergence of the correlation function of
deterministic processes with memory governed by $h(t)$ for any possible
values of $\theta$, we truncate the power law at some large time $T$, which we
call the ``integral time scale:''  it is the largest time scale up to which 
the memory of a past event survives. $T$ can thus be interpreted as the effective
Maxwell time of the relaxation process. The time dependence of $h(t)$
is an effective description of the relaxation of stress due to 
microscopic processes such as
dislocation motion, stress corrosion and hydrolytic weakening which obeys 
an Omori-like power law.

In summary, our model reads (in discretized form)
\be
\lambda(t) = \lambda_{\rm tec} ~ e^{\beta \omega(t)}~,~~~
\omega(t) = \sum_{i~|~t_i\leq t}  ~s(t_i) ~ h(t-t_i)
\label{vmjhi}
\ee
with the stress sources $s(t_i)$ distributed according to
a power law $P(s)$ with exponent $\mu$ and $h(t)$ having a power law memory.

We now derive our novel prediction for Omori's law quantifying the
decay of seismic activity after a ``mainshock'' occurring
at the origin of time. This amounts to determining
the typical time dependence of the seismic rate $\lambda(t)$ 
conditioned on a value $\lambda_M$
realized at $t=0$ which is larger than average. This formulation is due to the fact
that a mainshock of magnitude $M$ induces a local burst of seismic activity
proportional to $K~10^{\alpha M}$, where $K$ and $\alpha$ are two positive constants
\cite{Helm}. Since the stress sources are non-Gaussian but power law
distributed, their average and 
variance may not be defined. Rather than calculating the conditional expectation
of $\lambda(t)$, a typical measure of conditional seismicity rate can
be defined at any quantile level $q$ by the probability ${\rm Pr}[\lambda(t)>\lambda_q|\lambda_M]$
that the rate $\lambda(t)$ be larger than the quantile $\lambda_q$ conditioned on the
fact that the seismic rate was at some given value $\lambda_M$ at time $0$:
${\rm Pr}[\lambda(t)>\lambda_q|\lambda_M] = 
{\rm Pr}[e^{\beta \omega(t)}> {\lambda_q \over \lambda_{\rm tec}}|\omega_M]
={\rm Pr}[\omega(t)> (1/\beta)\ln \left({\lambda_q \over \lambda_{\rm tec}}\right)|\omega_M]$.
For Gaussian sources, $\omega$ is normally
distributed and we get
${\rm E}[e^{\omega(t)}|\omega_M] = \exp \left[ \beta {\rm E}[\omega(t)|\omega_M] 
+ {\beta^2 \over 2} {\rm Var}[\omega(t)|\omega_M] \right]$, 
where
${\rm E}[\omega(t)|\omega_M] = \omega_M~{{\rm Cov}[\omega(t),\omega_M] \over {\rm Var}[\omega_M]}$.
Using (\ref{vmjhi}), this would provide a closed formed expression for 
the Omori law describing the relaxation of the conditional rate ${\rm E}[\lambda(t)|\lambda_M]$.
The physical meaning of this result is that one can write a linear regression
$\omega(t) = \gamma(t) \omega_M + \epsilon$,
where $\gamma(t)$ is a non-random factor and $\epsilon$ is a centered Gaussian noise
with zero correlation with $\omega_M$. This equation writes that
the best predictor of $\omega$ given $\omega_M$ is $\gamma \omega_M$,
i.e., ${\rm E}[\omega(t)|\omega_M] = \gamma \omega_M$ with 
$\gamma = {{\rm Cov}[\omega(t),\omega_M] \over {\rm Var}[\omega_M]}$.
For power law stress sources, we use the 
insight that the natural generalization of the variance for power laws
$p(x) \approx C/x^{1+\mu}$ with infinite
variance (i.e., with $\mu<2$) is the scale parameter $C$ (see Chap.~4 of 
\cite{Sornette}). 
In the power law case, due to the linear form of $\omega$ in (\ref{vmjhi}),
we can still write $\omega(t) = \gamma(t) \omega_M + \epsilon$
but with $\omega(t), \omega_M$ and $\epsilon$
being power law distributed random variables with the same exponent $\mu$
and with scale factors equal respectively to $C_{\omega}$ (for
$\omega$ and $\omega_M$) and $C_{\epsilon}$.
The key idea is that $\gamma$ can be determined by forming the random variable defined as the 
product $\omega \omega_M = \gamma \omega_M^2 + \epsilon \omega_M$. It
is straightforward to show that the distribution of $\omega \omega_M$ consists of two
main contributions, (i) a dominant power law with exponent $\mu/2$ and scale
factor $C_{\omega \omega_M} = \gamma^{\mu/2}~C_{\omega}$, and (ii) a sub-dominant power law with 
exponent $\mu$ (with a logarithmic correction) and scale factor $C_{\omega}C_{\epsilon}$.
This has the following practical implication: if one measures or calculates
the leading power law decay of $\omega \times \omega_M$, the measure of its
scale factor gives access to the parameter $\gamma$ through the expression
$\gamma(t) = \left( C_{\omega \omega_M} / C_{\omega} \right)^{2 \over \mu}$.
where the time dependence of $\gamma(t)$ comes from that of $C_{\omega \omega_M}$.
For $\mu=2$, we recover the Gaussian result
with the correspondence $C_{\omega}= {\rm Var}[\omega]$ and 
$C_{\omega \omega_M}={\rm Cov}[\omega(t),\omega_M]$. 
Using (\ref{vmjhi}), we then form the product 
$\omega(t) \omega_M =  \sum_{i~|~t_i\leq t} \sum_{j~|~t_j\leq 0}  
~s(t_i)~s(t_j) ~ h(t-t_i)~h(-t_j)$, where the $s$'s are random variables 
with power law tail with exponent $\mu$.
Then, using standard calculations (see Chap.~4 of \cite{Sornette}), 
the terms in the double sum that contribute to the 
leading asymptotic power law tail with exponent $\mu/2$ correspond to the 
diagonal terms $i=j$, while all the other terms contribute to the sub-leading
power law tail with exponent $\mu$ with logarithmic corrections. This gives
the expression of the scale factor $C_{\omega \omega_M}^{\{\mu/2\}}$
of the dominating power law with exponent $\mu/2$ 
and finally yields
$\gamma = \left( \sum_{i~|~t_i\leq 0} 
\left[h(t-t_i) h(-t_i)\right]^{\mu \over 2} \right)^{2 \over \mu}$, in discrete form and 
\be
\gamma(t) = {c^{2(1+\theta)} \over \Delta t^{2/\mu}}  \left( {1 \over t^{2m-1}}
\int_{c/t}^{{T+c \over t}-1} dy {1 \over (y+1)^{m}}~
{1 \over y^{m}} \right)^{2 \over \mu}~,
\label{mgmlssaafa}
\ee
in continuous form where $m = (1+\theta)\mu/2$. The discrete time step $\Delta t$
converting the discrete into the continuous sum is the average time interval
between two events before a mainshock.

We thus obtain
${\rm Pr}[\omega(t)>y|\omega_M] = {\rm Pr}[\gamma \omega_M +\epsilon>y|\omega_M]
= {\rm Pr}[\epsilon>y-\gamma \omega_M|\omega_M] = {\bar F}(y-\gamma(t) \omega_M)$,
where ${\bar F}(\epsilon)$ is the complementary cumulative distribution of $\epsilon$.
Putting these results in (\ref{vmjhi}), this leads to
${\rm Pr}[\lambda(t)>\lambda_q|\lambda_M] = 
{\bar F}\left((1/\beta)\ln \left({\lambda_q \over 
\lambda_{\rm tec}}\right)-\gamma(t) \omega_M\right)$.
The typical time evolution of the seismicity rate $\lambda(t)$ conditioned
on the rate $\lambda_M$ at time $0$ is thus given by fixing the quantile probability
to some level ${\rm Pr}[\lambda(t)>\lambda_q|\lambda_M]=q$, leading to
\be
\lambda_q(t) = A_q ~\lambda_{\rm tec}~e^{\beta \gamma(t) \omega_M}~,
\label{gnnlblwd}
\ee
where $A_q = \exp \left( \beta {\bar F}^{-1}(q) \right)$.
The time-dependence of the seismic decay rate requires the determination of the time-dependence of 
$\gamma(t)$ given by (\ref{mgmlssaafa})). We now show that, for 
a rather broad range of values of the exponents
$\mu$ and $\theta$ defining the model, $\lambda_q(t)$ is approximately given by
\be
\lambda_q(t) \sim {1 \over t^{p(M)}}~,~~~p(M)= a \beta M + b \beta~,
\label{mgkhh}
\ee
where $a >0$ and $M$ is the mainshock magnitude. 

Consider first the case $2m=\mu (1+\theta) = 1$, such that
the exponent $m = (1+\theta)\mu/2$ defined in (\ref{mgmlssaafa})
equal to $1/2$. Then, ${d \gamma^{\mu/2} \over dt} =
 - {(h_0^2 /\Delta t^{2/\mu}) \over t}~\left[
{(T+c)^{1/2} \over (T+c-t)^{1/2}} - {c^{1/2} \over (t+c)^{1/2}}\right]$,
showing that ${d \gamma^{\mu/2} \over dt}$
is close to $-1/t$, and thus $\gamma^{\mu/2}(t)  \approx {\rm constant}_1 - 
{\rm constant}_2 \ln (t/T)$ which, for not too small nor too large $t$'s and 
for ${\rm constant}_1 < {\rm constant}_2$, gives
$\gamma^(t) \approx {\rm constant}'_1 - 
{\rm constant}'_2  \times \ln (t/T)$. This yields (\ref{mgkhh}).
Typically, the power law behavior is observed over more than two decades in time,
which is comparable to empirical observations, as verified by direct numerical
integration of (\ref{mgmlssaafa}). Then, expression (\ref{gnnlblwd})
leads to (\ref{mgkhh}) using the fact that 
$\omega_M \propto \ln (\lambda_M) \propto \ln (K~10^{\alpha M}) =\alpha \ln 10 ~M + \ln K$,
i.e., $\omega_M$ is linearly related to the magnitude $M$. 
The fact that $\gamma(t)$ is asymptotically exactly logarithmic in time
for $2m=\mu (1+\theta) = 1$ and thus that the seismic rate $\lambda(t)$ is
an Omori power law can be recovered from a different construction motivated 
by multiplicative cascades introduced in turbulence \cite{SchmittMarsan}.
This case covers the exact multifractal random walk model \cite{Muzy}, which 
corresponds asymtotically to $\theta=-1/2$ and $\mu=2$.
This continuous dependence of the exponent $p(M)$ has actually been documented
empirically in this case in another context of aftershock decay following shocks
in financial markets \cite{Sormu}.
For $2m=\mu (1+\theta) \neq 1$, one can often observe an 
approximate linear decay of $\gamma(t)$ as
a function of $\ln t$, over two to three order of magnitudes in time
in the decaying part, all the more so, the closer $m$ is to $1/2$, also 
leading to (\ref{mgkhh}).

We now show that this prediction is verified in the Southern Californian earthquakes
catalog with revised magnitudes (available from the Southern California 
Earthquake Center). The details of our analysis is given elsewhere \cite{longpaper}
and we summarize the main results. In order to improve the
statistical significance and to test for the stability of our analysis,
we analyzed four different sub-catalogs: 
$1932-2003$ for magnitude $M_L>3$ ($17,934$ events), 
$1975-2003$ for $M_L>2.5$ ($36,614$ events), 
$1992-2003$ for $M_L>2$ ($54,990$ events), and 
$1994-2003$ for $M_L>1.5$ ($86,228$ events).
We consider all events in a given sub-catalog and discriminate
between mainshocks and triggered events (``aftershocks''). Mainshocks are determined
by using two different declustering methods described below.
Once the mainshocks
are determined, triggered events are defined as those events
following a mainshock, which belong to a certain space-time neighborhood of it.
In order to test for the predicted dependence of the $p$-value as a function
of magnitude, we bin the mainshock magnitudes in intervals 
$[1.5; 2]$, $[2; 2.5]$, $[2.5; 3]$, and so on up to $[7; 7.5]$.
In each mainshock magnitude interval $[M_1; M_2]$, we consider
all triggered sequences emanating from mainshocks with magnitude in this interval 
and stacked them to a common origin of time. The resulting function is fitted 
using the modified Omori law $N(t)= B+ {a \over (t+c)^{p}}$,
\label{omorilaw}
where $B$ is a positive parameter introduced to account for the background seismicity
assumed to be superimposed over the genuine triggered sequences. The time shift $c$ 
ensures the regularization of the seismic rate at $t=0$. 

The first declustering method is essentially the same as defined in \cite{Helm}:
every event in the catalog is defined as a 
mainshock if it has not been preceded by an event with larger magnitude 
within a fixed space-time
window $T \times d$, with $T=1$ year and $d=50$ km. Looking for events
triggered by this mainshock, we define another space-time window following
it. The time dimension of the window is also set
to $1$ year, whereas the space dimension depends on the rupture length
of the main event. This spatial window is chosen as a circle of radius
equal to the mainshock rupture length $L=10^{-2.57+0.6M_L}$, which is
the average relationship
between $L$ and magnitude $M_L$ for California \cite{Wells}.
If any event falls within this space-time window, it is considered as
triggered by the main event. We have also checked the stability of the results by
considering a spatial neighborhood of radius $2L$ rather than $L$ for the 
triggered events. The second declustering method is
the same as the first one, except for
one element: the space window used for qualifying a mainshock is not fixed
to $d=50km$ but is chosen to adapt to the size of the rupture lengths $L(M_i)$ 
given by $L=10^{-2.57+0.6M_L}$ of all
events of all possible magnitudes $M_L(i)$ preceding this potential mainshock. 

\begin{figure}[h]
\includegraphics[width=8cm]{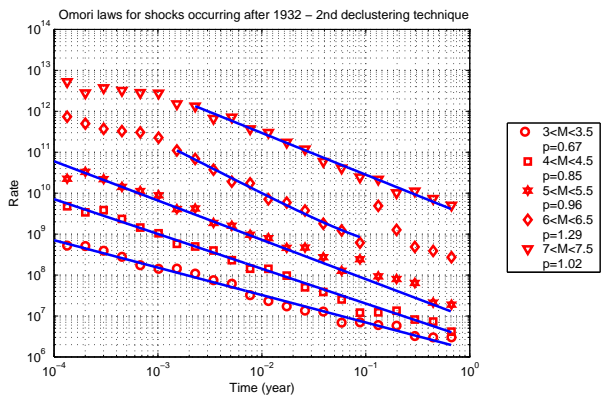}
\caption{\label{omoriplots3} Normalized seismic decay rates of stacked
sequences for several magnitude intervals of the mainshocks,
for the period from 1932 to 2003 when using
the second declustering technique.
}
\end{figure}

Figure \ref{omoriplots3} shows sets of typical seismic decay rates of stacked
sequences for several magnitude intervals of the mainshocks, 
for the period from 1932 to 2003 when using
the first declustering technique, with mainshock magnitudes above $M_L=1.5$.
Very similar plots are obtained for different time
periods, with the second declustering method and by varying 
the size from $L$ to $2L$ of the spatial domain
over which the triggered sequences are selected \cite{longpaper}. For large mainshock
magnitudes, the roll-off at small times is due to the observational
saturation and short-time lack of completeness of triggered sequences.

\begin{figure}[h]
\includegraphics[width=8cm]{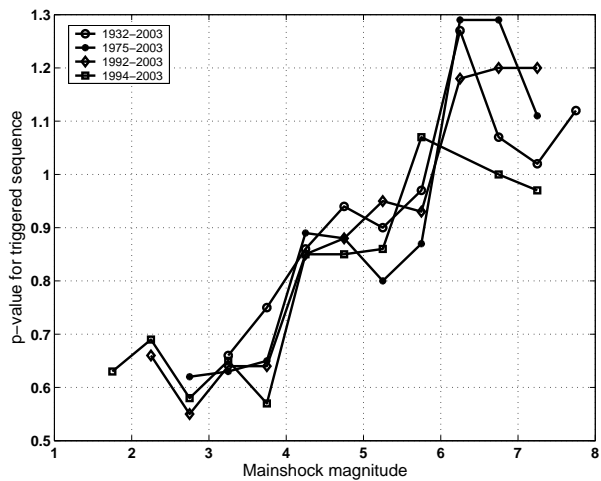}
\caption{\label{F3} $p$-values of the Omori law $\sim 1/t^p$
obtained by the procedure described in the text
for mainshocks (defined using the second declustering algorithm)
as a function of the main events' magnitude, for the different
sub-catalogs of lifespans given in the inset.
}
\end{figure}

Figure \ref{F3} shows the fitted
$p$-values as a function of the magnitude of the mainshocks for each of
the four sub-catalogs. 
We use a standard least-square fit of the seismic rate as a function of
time with a weight proportional to $t$ for each bin to balance their
relative importance. We also take into account the possible presence of a
background term. We have also performed maximum likelihood estimations 
of the exponent $p$, confirming the results shown in Fig.~\ref{F3} \cite{longpaper}.
To test the reliability and robustness 
of our results, we have simulated synthetic catalogs with the ETAS model
with known statistical properties following exactly the same
procedure as for the real catalogs. The ETAS model provides
a particular strong null hypothesis as it rationalizes most of the 
phenomenological statistical properties of earthquake catalogs \cite{Helmsor}.
By construction, synthetic catalogs generated with the ETAS model
should exhibit Omori laws with magnitude-independent exponents. Applying
our procedure to such synthetic catalogs
allows us to investigate whether the magnitude-dependence of the 
$p$-value reported above could result
from some bias introduced by our analysis rather than being 
a genuine property of earthquake catalogs. We verify that $p(M)$
obtained by our procedure is a constant independent of $M$ equal to the input value
used in the generation of the synthetic catalog \cite{longpaper}.

Let us conclude by offering an intuitive explanation of (\ref{mgkhh})
using the properties of multifractal spectra.
The temporal evolution of seismicity in a fixed spatial domain 
defines a statistically stationary measure
on the temporal axis, the measure determining the rate of earthquakes at any 
possible instant. 
An Omori sequence with exponent $p$ corresponds to a singularity (to the right) 
equal to $1-p$ (logarithmic for $p =1$). A large earthquake triggers 
a strong burst of seismicity, giving rise to a strong singularity. For the 
relation $\alpha = 1 - p$ to be consistent with the multifractal description,
a large earthquake must be associated with a strong singularity, a small $\alpha$,
hence a large $p$.
Reciprocally, small moment orders $q$ select weak seismic sequences, which are
thus associated with small local mainshocks. Small $q$'s are associated with large 
$\alpha$'s, hence small $p$'s.  By a similar argument in the space domain, 
the exponent of the spatial 
decay of the seismic rate induced by a mainshock of magnitude $M_L$ should
increase with $M_L$.  Thus, in this view, the ETAS model is nothing but the mono-fractal
approximation of a more general multifractal description of seismicity.

This work was partially supported by NSF-EAR02-30429 and by
the Southern California Earthquake Center (SCEC).

\end{document}